\documentclass{ws-p8-50x6-00}
\def\xgo{x_\gamma^{\rm OBS}}
\def\kt{{\rm k}_{\rm T}}
\def\met{\overline{E}_T}
\def\et2{\overline{E}_T^2}

\begin{document}

\title{Virtual Photon Structure at HERA}

\author{Dorian Kcira}

\address{ZEUS / DESY, Notkestr.85, 22607 Hamburg  Germany.\\E-mail: dorian.kcira@desy.de
\\$\;$\\
\rm On behalf of the ZEUS Collaboration.
}


\maketitle

\abstracts{
Measurements of the structure of the virtual photon in the
transition region between quasi-real photons and those far from
mass-shell have been made with the ZEUS detector at HERA, using an
integrated luminosity of $38$~pb$^{-1}$.  Dijet final states are
identified, and differential cross sections are presented in terms of
$\xgo$, an estimator of the fraction of the photon energy which
takes part in the QCD subprocess. Comparison is made to theoretical
predictions.
}

\section{Introduction}
\label{sec-int}

HERA allows for studies of jet production in a wide range of photon
virtualities, $Q^2$, and transverse energies of jets.
In the deep inelastic scattering (DIS) regime, $Q^2$ typically above $10$~GeV$^2$,
the interaction between the incoming positron and proton is mediated
by a structureless virtual photon.
On the other hand, in the photoproduction regime, $Q^2 \simeq 0$~GeV$^2$, the
photon may interact as an entity, i.e. the direct component, or as a source of
partons, i.e. the resolved component, where one of these partons, carrying a fraction
of the photon's momentum, $x_{\gamma}$, enters the hard subprocess.
Resolved photoproduction is commonly described
via parton distribution functions which receive contributions from both
perturbative and non-perturbative terms.

This paper studies jet production in a large range of photon virtualities,
including the transition region from photoproduction to DIS, with the
aim of understanding the evolution of the photon structure with increasing
photon virtuality.

The study of jet production introduces a second energy scale, e.g. $\met$,
the average transverse energy of the two highest $E_T$ jets in a given event,
and allows the determination of $x_{\gamma}$  through~\cite{Ze95}: 

\begin{equation}
\xgo = \frac{\sum_{\rm jets} E_T^{\rm jet} {\rm e}^{-  \eta^{\rm jet}} }{2 y E_e}\nonumber\;,
\end{equation}

where $E_T^{\rm jet}$ is the transverse energy of the jet and $\eta^{\rm jet}$ is the pseudorapidity,
defined as a function of the polar angle $\theta^{\rm jet}$:
$\eta^{\rm jet} = -\ln {(\tan{\displaystyle\frac{\theta^{\rm jet}}{2}})}$.

At leading order (LO) perturbative QCD for $\xgo > 0.75$, the direct component dominates,  while the $\xgo < 0.75$ region is sensitive mainly to the
resolved component. However, events with low values of $\xgo$ can also be produced when initial- and final-state
parton showers give rise to hadronic activity outside the dijet system. In pQCD, $\xgo$ is well defined
at all orders. Hence, measurements based on it can be compared with theoretical predictions at any
given order.

Triple differential cross sections with respect to $Q^2$, $\met$ and $\xgo$ are
compared to LO MC predictions based on different extrapolations of
the real photon structure to higher virtualities.

\section{Results and discussion} 

The dijet differential cross sections, $d^3\sigma  / ( d \xgo d Q^2 d \et2)$
corrected to the hadron level, have been measured using the $\kt$ jet algorithm in the
$\gamma^*p$ frame. The measured cross section is for dijet events in the pase-space
region defined by: $0.2 < y < 0.55$, and $0.1 < Q^2 < 0.55$~GeV$^2$, $1.5 < Q^2 < 10^4$~GeV$^2$.

\begin{figure}[t]
\epsfxsize=13pc 
\epsfbox{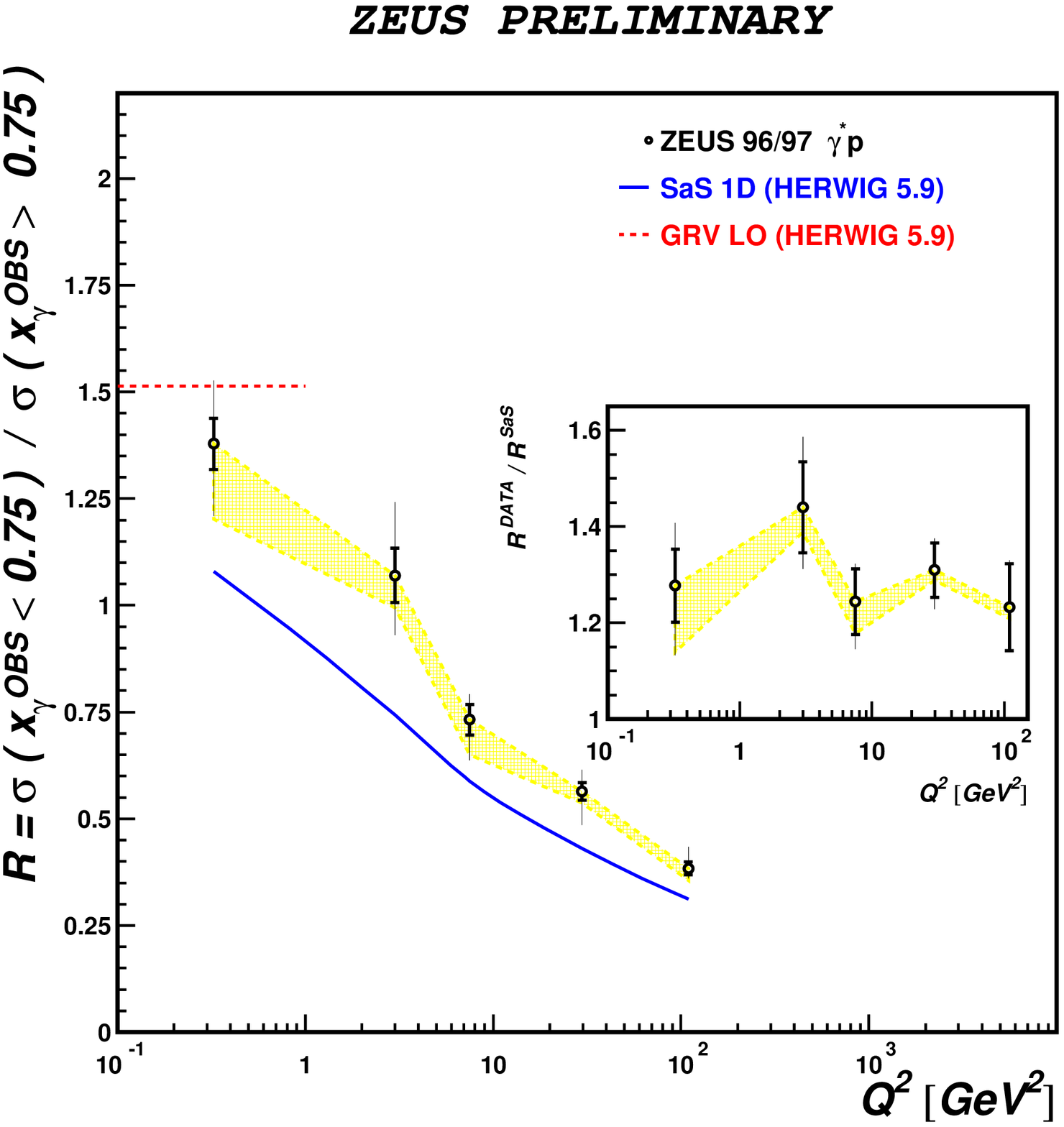} 
\epsfxsize=13pc %
\epsfbox{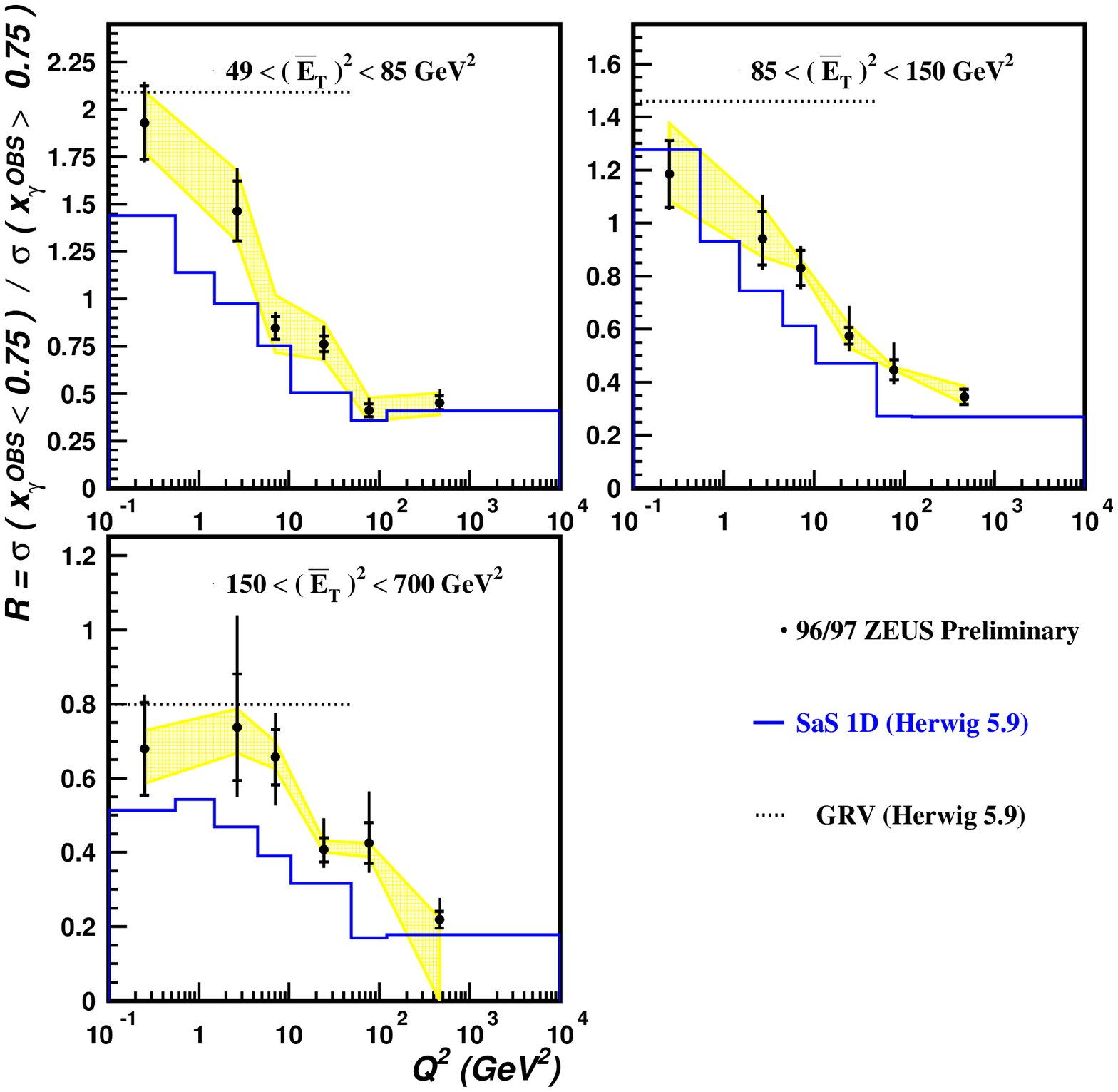} 
\caption{The ratio of cross sections $R = \sigma(\xgo<0.75)/\sigma(\xgo>0.75)$ as a function of $Q^2$ (left)
and the same ratio in different ranges of $\et2$ (right).
The solid line is the HERWIG MC using the SaS1D parametrization
for the photon PDFs and the dotted horizontal line is the HERWIG prediction using the GRV parametrization
for the real photon PDFs.
\label{fig:virtual_g4}}
\end{figure}

The ratio of cross sections $R = \sigma(\xgo<0.75)/\sigma(\xgo>0.75)$ as a function of $Q^2$ is shown
in Fig.~\ref{fig:virtual_g4} (left). The ratio falls with increasing $Q^2$. In the inset the ratio
of the data with the HERWIG~\cite{herwig6.2} prediction using the SaS1D parametrization~\cite{Sc96}
for the photon PDFs
is presented. There is no dependence of this ratio with the photon virtuality.
The ratio $R$  in three ranges of $\et2$ is also shown in Fig.~\ref{fig:virtual_g4} (right).
The ratio of the data falls with increasing $Q^2$ for each range of $\et2$.
These results can be interpreted as showing the suppression of the virtual photon structure with increasing photon
virtuality.
The HERWIG prediction using SaS1D also falls with increasing $Q^2$ but underestimates the measured
ratio.

\begin{figure}
\epsfxsize=13pc 
\epsfbox{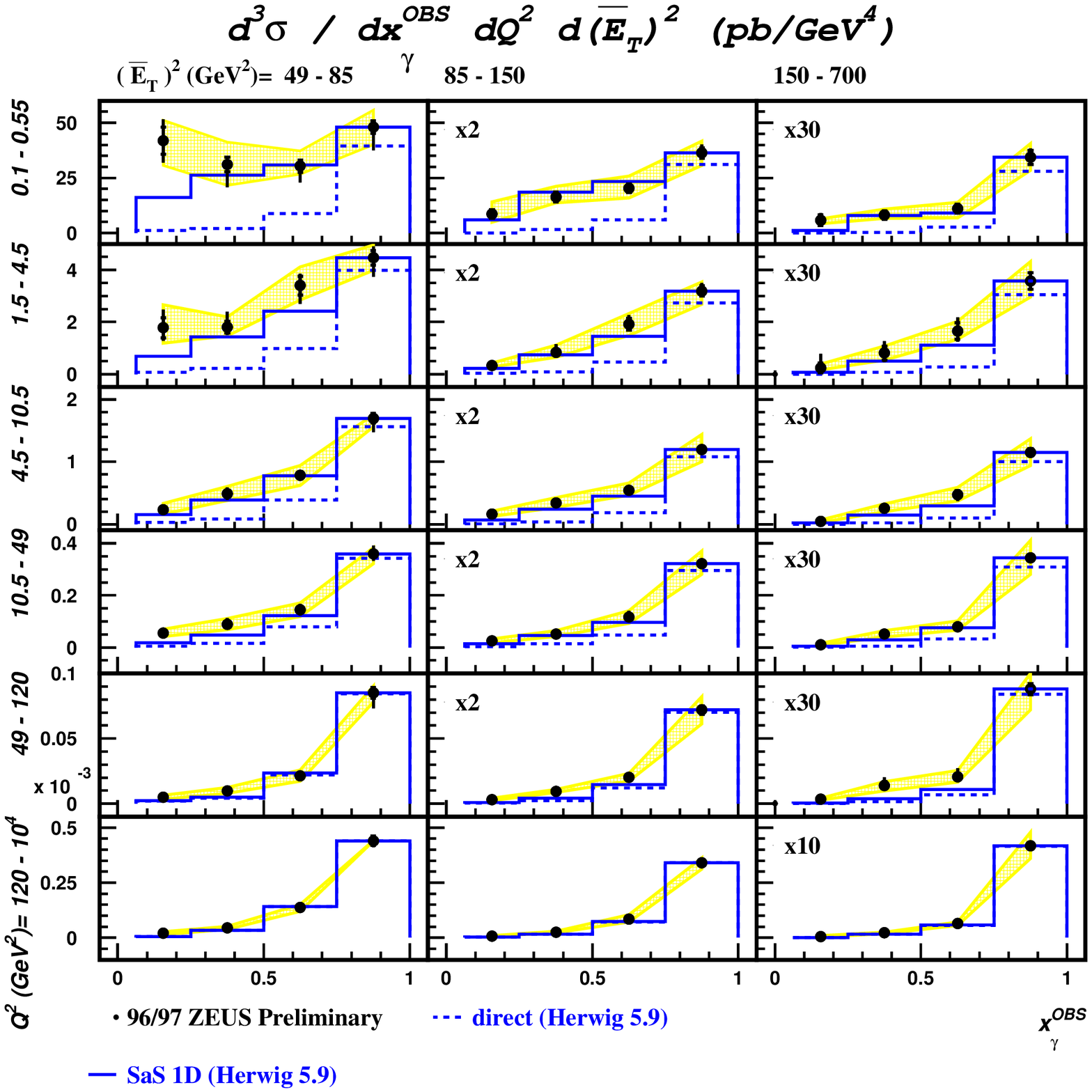}
\epsfxsize=13pc 
\epsfbox{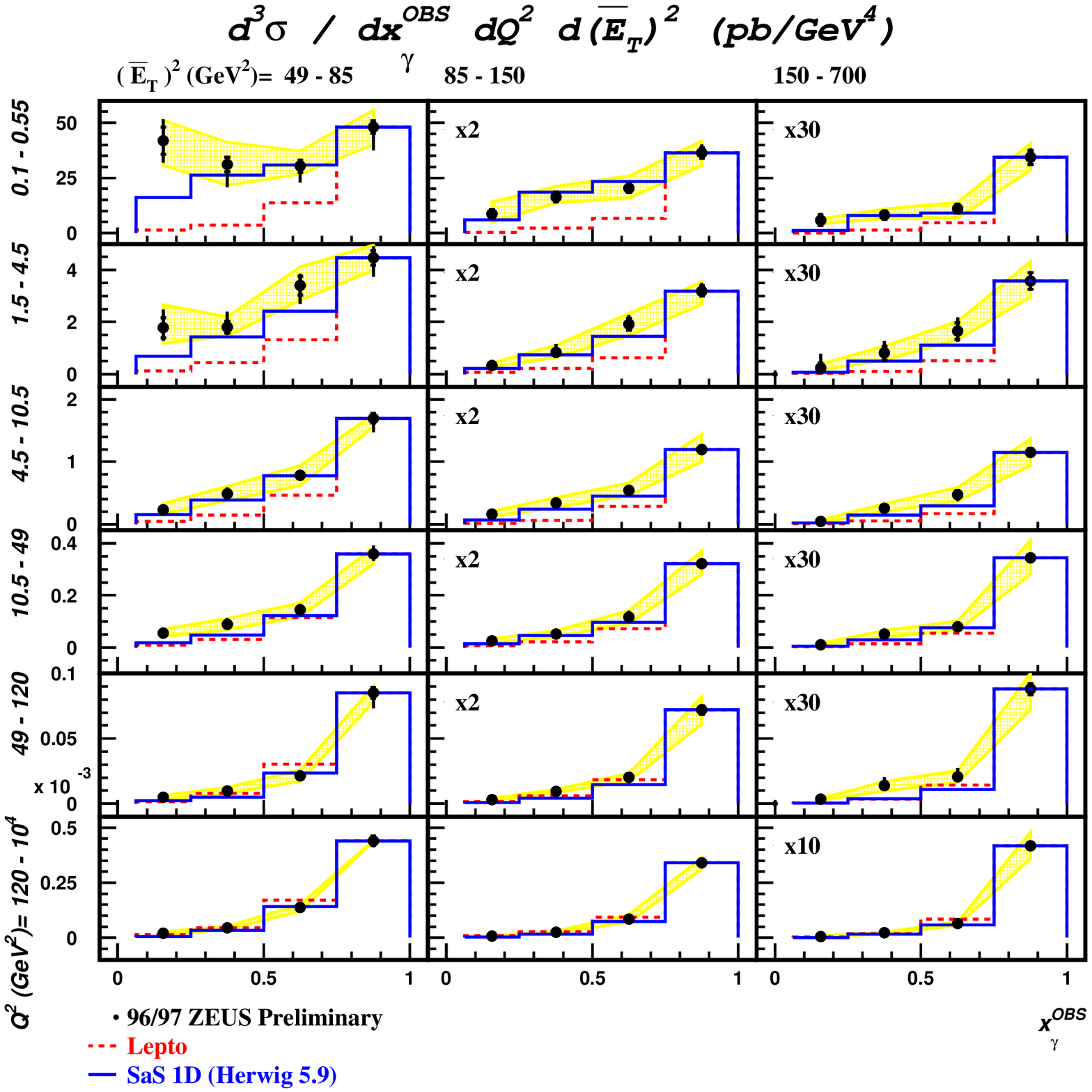}
\caption{Triple differential cross section $d^3\sigma  / ( d \xgo d Q^2 d \et2)$
as a function of $\xgo$ for different regions in $Q^2$ and $\et2$.
The points represent the measured cross sections with statistical errors (inner error bars) and uncorrelated
systematic errors added in quadrature to them (outer error bars).
The shaded bands display the uncertainty in the plotted quantities due to that in
the jet energy scale.
In the left plot the solid line is the HERWIG MC using the SaS1D parametrization for the photon PDFs
and the dashed line is the LO-direct part only from HERWIG. In the right plot the solid line is the
same and the dashed line is the prediction using the LEPTO MC.
\label{fig:virtual_g1}}
\end{figure} 

The measured triple differential dijet cross sections $d^3\sigma  / ( d \xgo d Q^2 d \et2)$
are shown as a function of $\xgo$ in Fig.~\ref{fig:virtual_g1} in
different regions of $Q^2$ and $\et2$.
In the left plot the LO HERWIG predictions using the SaS1D parametrization of the photon PDFs are shown
as well as the LO-direct part only of the HERWIG predictions. The HERWIG predictions
do not describe the absolute cross section of the
data and are therefore normalized to the highest $\xgo$ bin ($\xgo>0.75$) in order to compare
the shape of the data with that of the MC predictions.

For each $\et2$ bin, the cross section in the low $\xgo$ region falls faster with increasing
$Q^2$ than the cross section in the high $\xgo$ region.
For the bins with $Q^2>\et2$ the data are well described by the HERWIG predictions
including only the LO-direct component. In the bins with $Q^2<\et2$ the LO-direct
component is not enough to describe the data.

In the right plot of Fig.~\ref{fig:virtual_g1} a comparison of the cross sections with the LEPTO~\cite{lepto}
MC is shown.
For $Q^2>\et2$, LEPTO describes the data. LEPTO is a DIS MC generator and this
region is the one where LEPTO is expected to be valid.

\section{Conclusions}

Dijet triple differential cross sections,  $d^3\sigma  / ( d \xgo d Q^2 d \et2)$,
have been measured using the longitudinally invariant $\kt$ jet algorithm in the $\gamma^*p$
frame for $10^{-1} < Q^2 < 10^4$~GeV$^2$ and $0.2 < y < 0.55$. The cross sections are measured for jets with $ -3 < \eta^{\rm jet} < 0$,  $E_{T,1}^{\rm jet}>7.5$~GeV, and $E_{T,2}^{\rm jet}>6.5$~GeV in different bins of $Q^2$ and $\et2$.

The $\xgo$ dependence of the measured cross sections varies  with increasing $Q^2$ and $\et2$.
In each region of $\et2$, the low-$\xgo$ cross section decreases rapidly as $Q^2$ increases.

The predictions of HERWIG using the SaS1D photon PDFs describe the cross sections well for
$Q^2 > \et2$. In this region, the LO direct component alone describes the data well. For
$Q^2 < \et2$, a resolved component is needed.

The ratio $R = \sigma(\xgo<0.75) / \sigma(\xgo>0.75)$ for dijet cross sections
decreases as $Q^2$ increases in all the $\et2$ regions. However, the measured ratio
lies  above the predictions of HERWIG using the SaS~1D photon PDFs. The data may be interpreted
in terms of a resolved component that is suppressed as the photon virtuality increases.

\end{document}